\documentclass[conference]{IEEEtran}
\IEEEoverridecommandlockouts
\usepackage{cite}
\usepackage{amsmath,amssymb,amsfonts, dsfont}
\usepackage{algorithmic}
\usepackage{graphicx}
\usepackage{textcomp}
\usepackage{xcolor}
\def\BibTeX{{\rm B\kern-.05em{\sc i\kern-.025em b}\kern-.08em
		T\kern-.1667em\lower.7ex\hbox{E}\kern-.125emX}}

\begin{document}
	
\title{Robust Improvement of the Age of Information by Adaptive Packet Coding
		\thanks{Distribution A. Approved for public release: Distribution unlimited 88ABW-2020-3296 on Oct. 23, 2020.}
	}
	
\author{\IEEEauthorblockN{Maice Costa, Yalin E. Sagduyu, and Tugba Erpek}
		\IEEEauthorblockA{
			Intelligent Automation, Inc \\Rockville, MD 20855, USA \\
			\{mcosta, ysagduyu, terpek\}@i-a-i.com}
		\and
		\IEEEauthorblockN{Muriel M\'edard}
		\IEEEauthorblockA{
				RLE, Massachusetts Institute of Technology \\
		Cambridge, MA 02139, USA\\
		medard@mit.edu}
	}
	
\maketitle
	
\begin{abstract}
We consider a wireless communication network with an adaptive scheme to select the number of packets to be admitted and encoded for each transmission, and characterize the information timeliness. For a network of erasure channels and discrete time, we provide closed form expressions for the Average and Peak Age of Information (AoI) as functions of admission control and adaptive coding parameters, the feedback delay, and the maximum feasible end-to-end rate that depends on channel conditions and network topology. These new results guide the system design for robust improvements of the AoI when transmitting time sensitive information in the presence of topology and channel changes. We illustrate the benefits of using adaptive packet coding to improve information timeliness by characterizing the network performance with respect to the AoI along with its relationship to throughput (rate of successfully decoded packets at the destination) and per-packet delay. We show that significant AoI performance gains can be obtained in comparison to the uncoded case, and that these gains are robust to network variations as channel conditions and network topology change. 
\end{abstract}
	
	
\section{Introduction}
As wireless networks become more ubiquitous and diverse, different quality of service (QoS) requirements arise to serve different applications. In particular, many sensing, monitoring, decision, and control applications require the wireless network to be optimized to deliver timely information. Examples of such applications include surveillance systems, autonomous vehicles, healthcare monitoring, Internet of Things (IoT), and Ultra-Reliable Low-Latency Communication (URLLC) services in 5G. On the other hand, long-distance communications such as satellite communications and beyond line of sight (BLoS) high frequency (HF) communications impose long propagation delays that make the delivery of information in a timely manner even a more challenging task.
	
To address the need to design and optimize a wireless network to deliver fresh information about an observed process (such as a status update), metrics related to the Age of Information (AoI) have received much attention recently \cite{b0}. AoI provides means to quantify information timeliness (or freshness) when multiple observations are to be sequentially transmitted through a network to a destination that is interested in the most up-to-date observation, using metrics to characterize the average AoI \cite{b5} or the peak AoI \cite{b18}. 
	
Prior to this special attention paid to alternative performance metrics such as AoI, most work on network optimization has focused on the traditional metrics of throughput and delay. The benefits of packet level coding such as network coding to optimize the network throughput (beyond multi-hop routing) are well known \cite{b10,b11}. Capacity achieving schemes are available, promoting resiliency to channel erasures and reliability with more efficient feedback \cite{b8}. Network coding has been applied to optimize the throughput in a wireless network setting by considering channel effects, traffic dynamics, and interactions with other network layers \cite{b20, b12, b13, b21}. The importance of network coding for systems that require information timeliness has been highlighted in \cite{b9}, where the AoI is characterized in a multicast network with packet-level coding. The implication of timeliness for network coding design in terms of meeting hard delay deadlines has been studied in \cite{b16}.

While promoting improved throughput, network coding may incur other transmission and processing delay costs \cite{b14, b15, b17}, hence the characterization of the trade-off between delay sensitivity and throughput is crucial to reap the potential benefits of network coding. The trade-off can be formally captured using $\mathcal(l)_p$ norms of the packet arrival times, as shown in \cite{b2}. This class of delay metrics captures the average delay, and consequently the rate of transmission, at one extreme, and the maximum ordered inter-arrival delay at the other extreme. 
An adaptive coding scheme has been proposed in \cite{b1}, extending the generation-based random linear network coding \cite{b3} with a coding bucket of variable size, which takes the role of head-of-the-line, containing the packets that will be encoded together and sent through the wireless channel. The point-to-point scheme proposed in \cite{b1} has been extended in \cite{b2} to multi-hop line networks with several feedback schemes.

The proposed metrics based on $\mathcal(l)_p$ norms provide a framework to optimize the coding and scheduling depending on the desired level of delay sensitivity. However, when timeliness of information is the objective, optimizing the network with respect to throughput or delay may not be equivalent to optimizing the network with respect to AoI metrics.  In this work, we capture the delay-throughput trade-off through the characterization of AoI, and use the timeliness metrics to guide the design of a robust and efficient coding scheme. We highlight that the AoI encompasses the effects of throughput and delay, in addition to the effect of admission control in regulating the rate at which information is injected into the network. For example, a node may have outdated information about a process of interest if update messages are not available to be injected in the network. In this case, even if the delays are small, the AoI at the destination node would be large. Hence, characterizing the AoI is important to optimize the system when the application requires timely information.

In this paper, we study the intricate relationships between AoI, per-packet delay, and throughput metrics, and demonstrate that adaptive network coding promotes a significant AoI performance improvement that is robust with respect to variations of channel conditions and network topology. In a network of erasure channels, packets are combined using random linear network coding (RLNC) at the source and sent towards a destination that is interested in fresh information of decoded packets and sends ACKs reaching back the source after a feedback delay. The source can control the packet admission and the number of packets to be encoded in each transmission for timely information delivery. In this setting, our contribution is to provide closed form expressions for the Average and Peak AoI as functions of the design parameters for the admission control and adaptive coding schemes, the feedback delay, and the maximum rate that can be achieved from the source to the destination through different networks, ranging from single-hop to multi-hop and multi-path. We demonstrate that adaptive coding sustains a robust improvement in the presence of changes in channel conditions and network topology.

The rest of the paper is organized as follows. Section II describes the adaptive coding scheme that we consider in this paper. Section III presents the timeliness performance metrics. Section IV analyzes performance trade-offs for both point-to-point communications and multi-hop line networks. Section V presents numerical results. Section VI concludes the paper. 

\section{Adaptive Coding}

\begin{figure}[t]
\centerline{\includegraphics[width = 1\columnwidth]{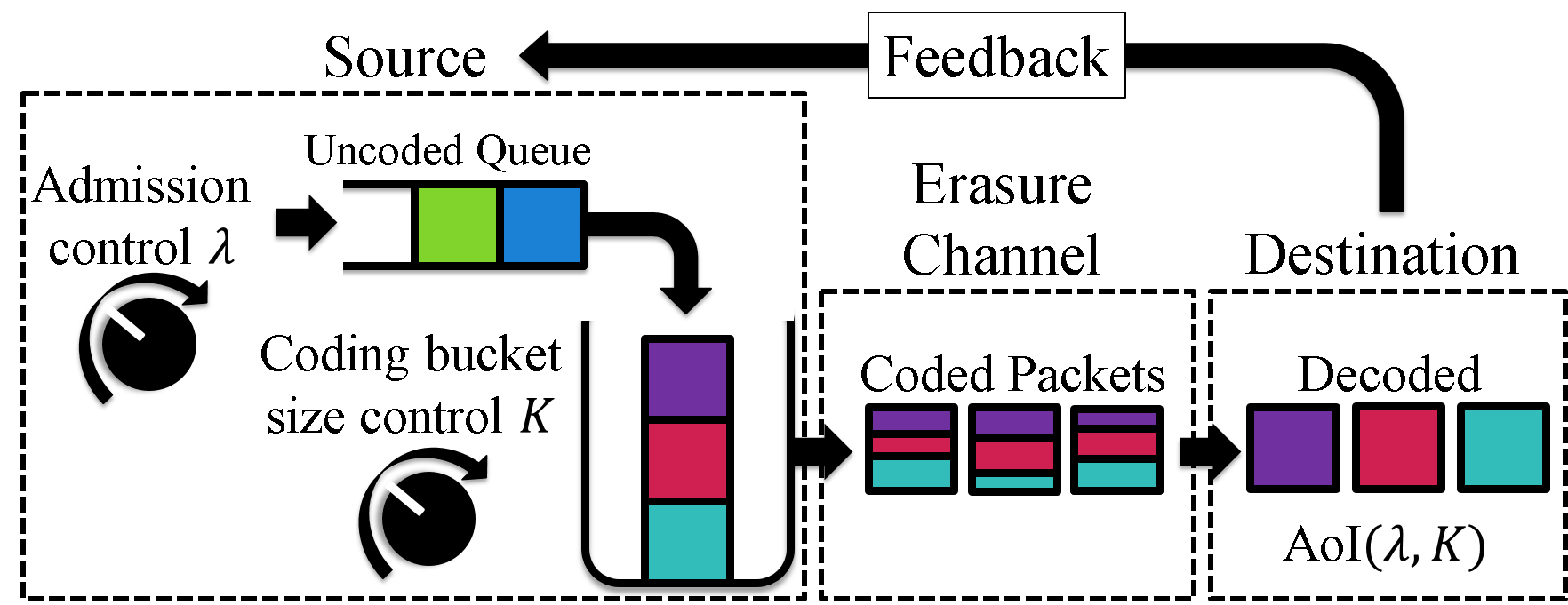}}
\caption{Model: adaptive network coding is performed at source node, coded packets are transmitted through an erasure channel, feedback is sent from destination when packets are decoded.}
\label{fig:Model}
\end{figure}

We consider the adaptive coding scheme proposed in \cite{b1} to transmit status update packets through a network of erasure channels. Adaptive coding is illustrated in Fig.\ref{fig:Model}. The source may apply admission control to regulate the packet arrival rate $\lambda$. An example of a simple admission control is to sample from the arriving process, keeping packets with probability $\beta$ and discarding packets with probability $1-\beta$. The admitted packets are kept in a buffer, in order of arrival. The source node keeps a coding bucket with up to $K$ packets, corresponding to the \textit{head of the line} used in generation-based RLNC \cite{b3}. 

The packets in the bucket are coded using RLNC. A coded packet is a linear combination of all the packets in the bucket, with a vector of coefficients drawn uniformly at random from the space $\mathds{F}_q^K$ over a finite field $\mathds{F}_q$ of size $q$. We assume that the field is sufficiently large, so that coded packets are linearly independent, and every packet successfully received at the destination is informative. Coded packets are generated and transmitted, one per time slot, as long as there is at least one packet in the bucket. The maximum feasible rate $r$ that can be achieved from the source to the destination depends on erasure probabilities and network topology. For a single link, $r=1-\epsilon$ for erasure probability $\epsilon$. We determine $r$ for multi-hop and multi-path networks in Section \ref{sec:topo}. An arriving packet will be included in the next coded packet if upon arrival it finds the bucket with less than $K$ packets, otherwise it will wait in the queue. The source node transmits linear combinations of the packets in the coding bucket, together with the coding coefficients (note that this overhead is negligible compared to the packet size), until the destination node receives enough packets to decode them all using Gaussian elimination.   

The destination sends an ACK message to the source, and we assume this feedback arrives successfully after the delay $D$. Once the feedback is received, the source node empties the coding bucket by discarding those packets, and moves new packets that may be waiting in the queue. The maximum number of packets included in the coding bucket, $K$ can be determined depending on timeliness and throughput requirements. The throughput, denoted with $\mu$, is defined as the rate of packets that are successfully decoded at the destination, and is a function of $r$, $D$ and $K$, as well as the packet size $L$, as we discuss in more detail in Section \ref{sec:tradeoffs}. We characterize the trade-off between performance metrics related to timeliness and the more traditional metrics of throughput and delay. 

\section{Timeliness Performance Metrics}

We use the AoI metrics to measure information timeliness. AoI is a time-evolving process evaluated at the destination that tracks the time since the last received update was generated. Let $U(t)$ be the time at which the most up-to-date message was generated. For discrete (slotted) time, the AoI evolves as
\begin{equation}
A(t+1) = 
\begin{cases}
      A(t)+1 & \text{if no update received,}\\
      \min\{t-U(t),A(t)\} & \text{if update received.}
    \end{cases}  
\label{eq:discreteAoI}
\end{equation}

The commonly used metrics to characterize the AoI are the Average AoI (AAoI), which we denote with $A_A$, and the Peak AoI (PAoI), which we denote with $A_P$. For the case of discrete time, these metrics can be formally defined as
\begin{equation}
A_A \triangleq \limsup_{T\rightarrow\infty}\frac{1}{T}\sum_{t=1}^{T}A(t),
\label{eq:AAoI}
\end{equation}

\begin{equation}
A_P \triangleq \limsup_{T\rightarrow\infty}\frac{\sum_{t=1}^{T}A(t)\mathds{1}_{\{A(t+1)\leq A(t)\}}}{\sum_{t=1}^{T}\mathds{1}_{\{A(t+1)\leq A(t)\}}}.
\label{eq:PAoI}
\end{equation}

Assuming a Bernoulli arrival process of rate $\lambda$ at the source node, a general service time distribution, and a single server (or a single path from source to destination), we have a Ber/G/1 queue. Let $S$ denote the service time, with $\mathbb{E}[S]=1/\mu$ and $\rho=\lambda\mathbb{E}[S]$. The waiting time in the Ber/G/1 queue \cite{b6} is given by 
\begin{equation}
\mathbb{E}[W]=\frac{\lambda\mathbb{E}[S^2]-\rho}{2(1-\rho)}.
\label{eq:wait}
\end{equation}

The AAoI and PAoI for the Ber/G/1 queue have been presented in \cite{b7} and are given by
\begin{equation}
A_A = 1+\frac{1}{\mu}+\frac{(1-\lambda)(1-\rho)}{\lambda \mathcal{L}_S(1-\lambda)}+\frac{\lambda \mathbb{E}[S^2]-\rho}{2(1-\rho)}, \;\text{and}
\label{eq:AAGeo}
\end{equation}
\begin{equation}
A_P = \frac{1}{\lambda}+\frac{1}{\mu}+\frac{\lambda \mathbb{E}[S^2]-\rho}{2(1-\rho)},
\label{eq:APGeo}
\end{equation}
where $ \mathcal{L}_S(\cdot)$ represents the probability generating function (PGF) of the service time $S$. 

\section{Performance Trade-Offs}\label{sec:tradeoffs}
\subsection{Point-to-Point Communication}
We denote with $L$ the packet length and with $N$ the number of original packets that should be delivered. We also let $\Delta T_i$ represent the ordered inter-delivery time for the $i$th packet. That is, if $T_i$ represents the in-order delivery time of the $i$th packet, then $\Delta T_i = T_i - T_{i-1}, \; i\in\{1,\ldots,N\}$, where we define $T_0\triangleq-D$ to account for the feedback delay $D$ associated to each bucket of at least one packet. The delay sensitivity of the receiver is modeled using the $\mathcal{L}_p$-norm of the sequence of variables $(\Delta T_1, \Delta T_2,\ldots,\Delta T_N)$ with a delay cost function defined by \cite{b1}:
\begin{equation}
d(p)=\frac{1}{L} \left[\frac{1}{N} \sum_{i=1}^{N} \mathbb{E}[\Delta T_i]^p \right]^{\frac{1}{p}} ,\;\;\;p\in[1,\infty).
\label{eq:lpnorm}
\end{equation}                     
When $p=1$, we have the average delay per packet, normalized by the total size of the received data. Hence, minimizing $d(1)$ is tantamount to maximizing the throughput $\mu= d(1)^{-1}$.
When $p=\infty$, we have $d(\infty)=\frac{1}{L}\max_{i=1\ldots N} \Delta T_i$, which is the maximum expected inter-arrival time between two successive packets, or per-packet delay. Hence, minimizing $d(\infty)$ is equivalent to minimizing the per-packet delay.

Consider the transmission of $N$ packets using coding buckets with $K$ packets.  We identify the packets in the $i$th bucket as $\{P_{i1}, \ldots,P_{iK}\}$, and denote with $\Delta T_{ij}$ the inter-delivery time of the $j$th packet within the $i$th bucket. The cost function in \eqref{eq:lpnorm} can be rewritten as
\begin{equation}
d(p)=\frac{1}{L} \left[\frac{1}{N}\frac{N}{K}\sum_{j=1}^{K} \mathbb{E}[\Delta T_{ij}]^p \right]^{\frac{1}{p}} ,\;\;\;p\in[1,\infty).
\label{eq:lpnorm2}
\end{equation}      

Note that, within the same bucket, the packets are assumed to have the same delivery time, which is the time when all the packets can be decoded at the destination node. Let $X$ represent the time to deliver the entire bucket, while $r$ denotes the maximum feasible end-to-end rate. Then, $\Delta T_{ij}=X+D$ for $j\equiv1 $ (mod $K$), and zero otherwise. Under the assumption of an erasure channel, the number of time slots $X$ needed to successfully deliver $K$ linearly independent coded packets follows a negative binomial distribution $NegBin(r,K)$, with $\mathbb{E}[X]=K/r$.  Using these properties in \eqref{eq:lpnorm2}, yields
\begin{equation}
d(p)=\frac{1}{LK^{\frac{1}{p}}}  \left(\frac{K}{r}+D\right).
\label{eq:dp}
\end{equation} 

If we look at a single packet, as opposed to the entire bucket, the service rate is $S=\frac{X+D}{LK}$, where $X$ is the random variable representing the time to deliver all the $K$ packets in a bucket, $D$ is a constant feedback delay, $L$ is the packet length, and $K$ is the bucket size. As a result, we have
\begin{eqnarray}
\mathbb{E}[S] & = & \frac{1}{rL}+\frac{D}{LK},
\\
Var(S) & = & \frac{1-r}{L^2K r^2},
\\
\mathbb{E}[S^2]&=& \frac{(1-r)K+(K+Dr)^2}{(LrK)^2},
\\
\mathcal{L}_S(z)&=&z^{\frac{D}{LK}}\mathcal{L}_X(z^{\frac{1}{LK}}).
\end{eqnarray}

Note that $\mathbb{E}[S]=d(1)$. Also, for a negative binomial random variable $X$ representing the number of attempts until $K$ successful transmissions, we have 
\begin{equation}
\mathcal{L}_X(z)=\left(\frac{rz}{1-(1-r)z}\right)^K, 
\label{eq:Lx}
\end{equation}
hence for the service time of each packet, the PGF is 
\begin{equation}
\mathcal{L}_S(z)=z^{\frac{D}{LK}}\left(\frac{rz^{\frac{1}{LK}}}{1-(1-r)z^{\frac{1}{LK}}}\right)^K.
\label{eq:LS}
\end{equation}

Using \eqref{eq:dp}--\eqref{eq:LS} together with \eqref{eq:AAGeo} and \eqref{eq:APGeo} yields closed form expressions for AAoI and PAoI as a function of $\lambda$, $L$, $D$, $r$, and $K$. For the PAoI, we obtain
\begin{IEEEeqnarray}{rCl}
A_P &=& \frac{1}{\lambda}+\frac{K+Dr}{KLr}+ \nonumber\\
&&\frac{\lambda\left[K(1-r)-(KLr)(K+Dr)+(K+Dr)^2\right]}{2[(KLr)^2-\lambda(KLr)(K+Dr)]},\IEEEeqnarraynumspace
\label{eq:APp1}
\end{IEEEeqnarray}
which can be used together with \eqref{eq:LS} to obtain the expression for the AAoI as
\begin{IEEEeqnarray}{rCl}
A_A &=&A_P+\left(1-\frac{1}{\lambda}\right) + \nonumber\\
&&\frac{(1-\lambda)\left(1-\frac{\lambda(K+Dr)}{KLr}\right)}{\lambda (1-\lambda)^{\frac{D}{LK}}\left(\frac{r(1-\lambda)^{\frac{1}{LK}}}{1-(1-r)(1-\lambda)^{\frac{1}{LK}}}\right)^K }.\IEEEeqnarraynumspace
\label{eq:AAp1}
\end{IEEEeqnarray}

Note that the variables that can be controlled are the arrival rate $\lambda$, which can be modified through admission control policies, and the bucket size $K$, which can be adapted according to the sensitivity to the timeliness, as well as the delay and throughput constraints. From \cite{b1}, we have the mean maximum inter-delivery time $d (\infty) = \frac{K}{rL} + \frac{D}{L}$, which increases with $K$, and the throughput given by $d(1)^{-1}$ benefits from larger $K$. On the other hand, $d(1)= \frac{1}{rL}+\frac{D}{LK}$ decreases with $K$. The metrics related to timeliness capture this trade-off between throughput and delay, and provide guidance to select the optimal values of $K$. We evaluate these trade-offs numerically in Section \ref{sec:perf}.

\subsection{Robustness to Different Network Topologies}\label{sec:topo}
The previous results hold for different network topologies, under the assumption that the propagation delay is negligible compared to the delays caused by network congestion and by the need to wait for sufficient degrees of freedom in order to decode the packets. We assume that the number of packets in a bucket remains the same at each intermediate node. At each hop, the node mixes the packets it has together and sends it. The feedback is an end-to-end feedback as described in Fig.~\ref{fig:Model}. 

To extend to a multi-hop network, we consider a network consisting of $H$ links in tandem, each link with erasure probability $\epsilon_h$, $h = 1,\ldots,H$, as described in \cite{b2}. Given our focus on systems where the timeliness of information is relevant, we consider that intermediate nodes do not attempt to decode packets, and perform a recode-and-forward scheme. In this case, the end-to-end rate at which the encoded packets are received at the destination is obtained through the min-cut max-flow theorem as
\begin{equation}
r_{\text{multihop}} = \min_{h=1,\ldots,H}(1-\epsilon_h).
\label{eq:mh_rate}
\end{equation}

For the multi-path case, the main challenge is to determine the allocation of the new coded packets of information to be transmitted over the available paths. It is necessary to balance between the number of packets, sent to maximize throughput, and the number of retransmissions due to channel erasures, sent to minimize delay. This problem was addressed in \cite {b19}, where the authors proposed a bit-filling scheme to allocate transmitted packets to the paths at each time slot. Consider the availability of $Z$ paths, with independent erasure probabilities given by $\epsilon_j$, where $j\in\{1,\ldots,Z\}$. Let the number of packets allocated to path $j$ be denoted with $k_j$, such that all packets in the coding bucket are allocated to a path, satisfying the sum constraint $\sum_{j=1}^{Z}z_j=K$.
The delivery time is determined by the path with maximum number of transmissions, so we write
\begin{equation}
T_{\text{multipath}} = \max_{j=1, \ldots Z} \left(k_j\frac{1}{1-\epsilon_j}\right).
\label{eq:mp_rate}
\end{equation}
The minimization of $T_{\text{multipath}}$ subject to the sum constraint is solved using discrete water filling, and the resulting end-to-end rate is $r_{\text{multipath}}=K/T_{\text{multipath}}$.

\begin{figure}[tbp]
\centerline{\includegraphics[width=0.9\columnwidth]{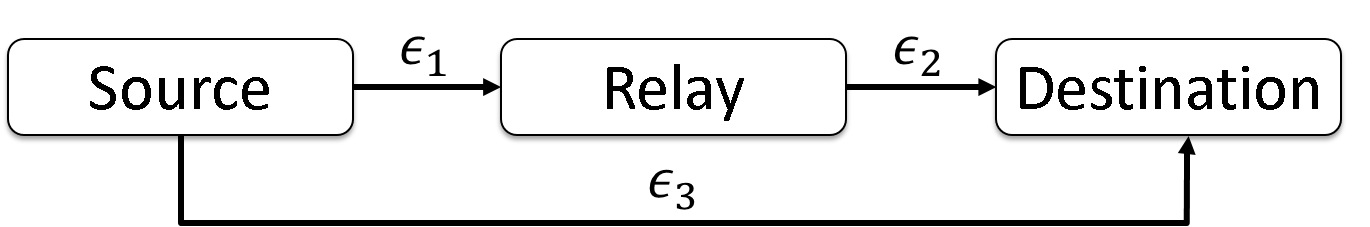}}
\caption{Simple network topology for multi-hop multi-path illustration.}
\label{fig:3node}
\end{figure}

For illustration purposes, consider the simple topology in Fig.~\ref{fig:3node}. In this case, there is a multi-hop path that we identify as the \textit{relayed} path, and there is a \textit{direct} path to destination. The discrete water filling will allocate $k_{\text{relayed}}$ and $k_{\text{direct}}$ to each path, such that $k_{\text{relayed}}+k_{\text{direct}}=K$. The resulting end-to-end rate in this particular scenario becomes
\begin{equation}
r_{\text{multipath}}=\frac{K}{\max\left(\frac{k_{\text{relayed}}}{\min((1-\epsilon_1),(1-\epsilon_2))},\frac{k_{\text{direct}}}{(1-\epsilon_3)}\right)}.
\label{eq:mod_rate}
\end{equation}

Effectively, the multi-hop and multi-path scenarios change the maximum feasible end-to-end rate we denoted with $r$, and our previous results hold when replacing $r$ with the correct representation as discussed above. Processing times in each hop would simply shift the results. Hence, the results we present hold for different network topologies (refer to Fig.~\ref{fig:AgeRate} and Table~\ref{tab:gains} for numerical examples).

\section{Performance Evaluation} \label{sec:perf}

\begin{figure}[tbp]
\centerline{\includegraphics[width=0.8\columnwidth]{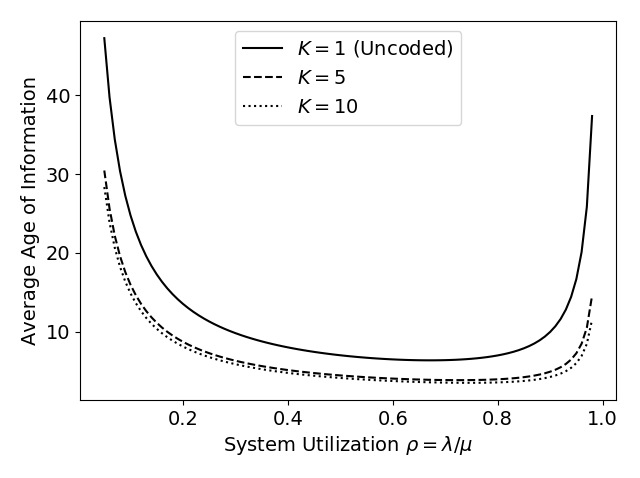}}
\vspace{-0.3cm}
\caption{AAoI versus system utilization factor $\rho=\lambda/\mu$.}
\label{fig:AgeRho}
\end{figure}

Unless specified otherwise, we set the packet length $L=1$, the feedback delay $D=1$, the channel utilization factor $\rho=0.6$, and the rate $r=0.8$. The arrival rate is limited by the service rate to guarantee the stability of the queue, which is tantamount to limiting the system utilization factor $\rho\triangleq \lambda/\mu<1$. Note that $\mu=d(1)^{-1}$ in this case. In general, varying the arrival rate $\lambda$ produces two effects in AoI metrics. While increasing the arrival rate may result in delivering messages more often to the destination, reducing the AoI, it may also result in network congestion and larger delays, increasing the AoI. The two effects are observed when plotting the AoI versus $\rho$, which also describes the arrival rate as a fraction of the service rate, as shown in Fig.~\ref{fig:AgeRho}. We note that departing from the uncoded case $K=1$ provides the most significant gains with respect to the AoI metrics.

\begin{figure}[tbp]
\centerline{\includegraphics[width=0.8\columnwidth]{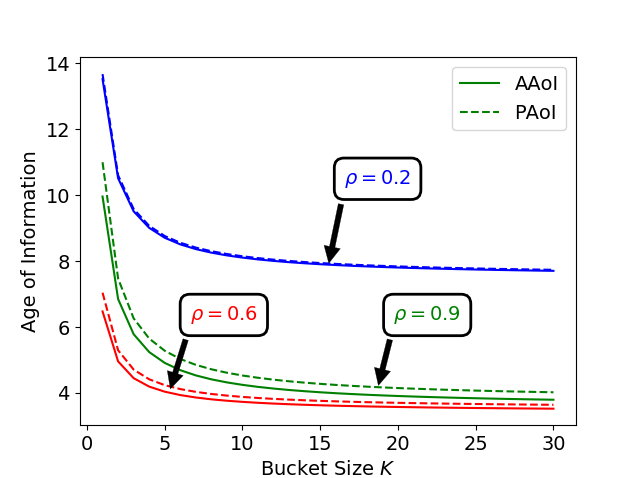}}
\vspace{-0.2cm}
\caption{AoI versus coding bucket size $K$ for different levels of congestion represented by $\rho$.}
\label{fig:AgeK}
\end{figure}

Fig.~\ref{fig:AgeK} shows the effect of the coding bucket size $K$, and presents curves for different values of the system utilization factor $\rho$, illustrating the trade-off between the metrics of AoI, delay, and throughput. Clearly, the AoI is decreasing with $K$ and, once more, we note that departing from the uncoded case $K=1$ provides significant performance gains with respect to timeliness. While the scenarios with more congestion (larger $\rho$) can benefit from further increasing $K$, we note that, in general, it is sufficient to combine a very small number (such as $10$) of original packets into a coded packet in order to obtain significant improvement in timeliness for all levels of congestion. We note that, for any given value of $K$, there is significant gain in increasing the utilization factor from $\rho=0.2$ to $\rho=0.6$. In this range, we can increase throughput, and the destination can receive updated messages more often. Further increasing the utilization factor from $\rho=0.6$ to $\rho=0.9$ creates congestion, increasing the waiting times in queue, and increasing the AoI. For larger $\rho$ the PAoI metric also becomes more distinct from the AAoI and can be used as an upper bound on AoI, given that it is simpler to calculate. 

\begin{figure}[tbp]
\centerline{\includegraphics[width=0.8\columnwidth]{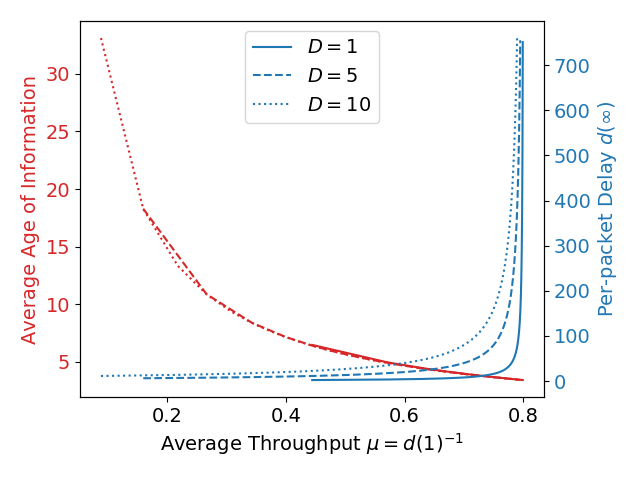}}
\vspace{-0.2cm}
\caption{AAoI $A_A$ and per-packet delay $d(\infty)$  versus throughput $\mu=d(1)^{-1}$. Channel utilization is fixed to $\rho=0.6$.}
\label{fig:Aged1}
\end{figure}

Fig.~\ref{fig:Aged1} depicts, on the left, the AoI versus throughput $\mu = d(1)^{-1}$. We obtain the curves by changing the bucket size $K\in [1,\infty)$. Increasing $K$ yields larger throughput and reduces the AoI, even though for very large $K$ the maximum expected delay for a packet may increase significantly, as it takes longer to decode all the packets in the bucket. This trade-off is illustrated with the curves associated to the axis on the right, which shows $d(\infty)$ as a function of the throughput. Under the conditions assumed in this case, the AoI is dominated by $d(1)$. The effect of the feedback delay is also represented in Fig.~\ref{fig:Aged1}. It determines the range of feasible pairs $(d(1),d(\infty))$ and also impacts the acceptable range of arrival rates such that the system is stable. Larger $D$ requires smaller arrival rates, and also results in larger average delays $d(1)$, which result in significant increase in AoI. 

\begin{figure}[tbp]
\centerline{\includegraphics[width=0.8\columnwidth]{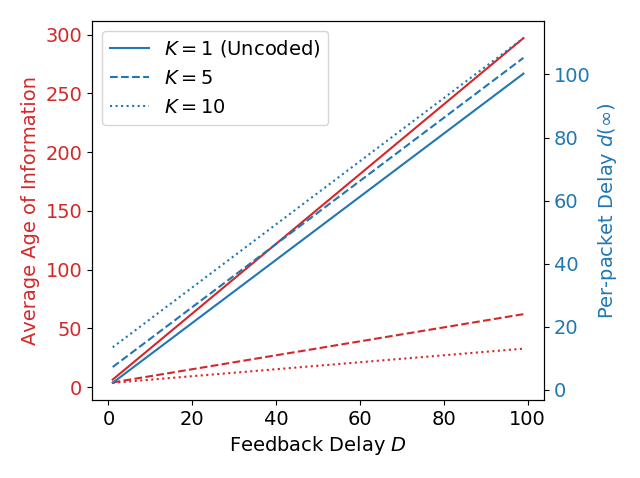}}
\caption{AAoI $A_A$ and per-packet delay $d(\infty)$ versus feedback delay $D$. Channel utilization is fixed to $\rho=0.6$.}
\vspace{-0.4cm}
\label{fig:AgeD}
\end{figure}

We illustrate the effect of the feedback delay $D$ in Fig.~\ref{fig:AgeD}. Both the AoI and the delay increase with $D$, as expected. However, the impact of $D$ on the AoI decreases with $K$, and a larger $D$ can be tolerated by using a larger $K$. Meanwhile, the per-packet delay $d(\infty)$ presents similar slope, i.e. increases with $D$ at similar rate, for different values of $K$.

\begin{figure}[tbp]
\centerline{\includegraphics[width= 0.8\columnwidth]{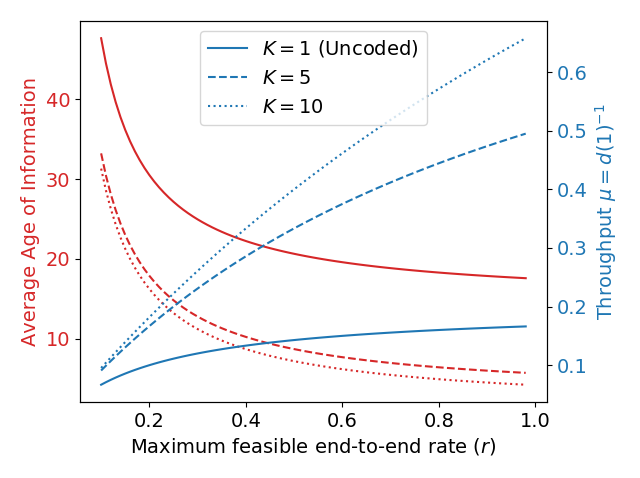}}
\vspace{-0.3cm}
\caption{AAoI $A_A$ and throughput $\mu=d(1)^{-1}$ versus maximum feasible end-to-end rate $r$. Channel utilization $\rho=06$. and feedback delay $D=5$.}
\label{fig:AgeRate}
\end{figure}

Fig.~\ref{fig:AgeRate} shows the AoI and throughput as a function of $r$. For single hop, $r=1-\epsilon$, while for multi-hop and multi-path we can use \eqref{eq:mh_rate} and \eqref{eq:mp_rate} to obtain a modified rate as exemplified by \eqref{eq:mod_rate}. By modeling a multi-hop multi-path network using a modified erasure channel, we note that the performance improvement with respect to timeliness is robust to the network topology. Nonetheless, the performance is affected by the maximum end-to-end rate that can be achieved in the network. In the case of AoI, the rate has more impact for $r<0.2$. In that range, an increase in rate results in significant reduction of the AoI. In other words, except in the case of very poor channel conditions, the performance with respect to AoI is very stable to variations in the rate. In the case of throughput, the performance improves more steadily with the increasing rate. For both AoI and throughput we observe significant gains in departing from the uncoded case ($K=1$). We highlight that the gain of using $K>1$ is observed under any channel conditions. The reduction in AoI values with respect to the uncoded case increases as the channel conditions deteriorate, and it is significantly larger for larger feedback delay, as shown in Table~\ref{tab:gains}, indicating that the adaptive coding scheme provides a robust improvement of timeliness, even under unfavorable channel or network conditions.  

In summary, we observed that the adaptive coding scheme is very robust to variations in network topology, channel conditions, and size of coded packets. It requires only a small code to produce significant performance improvement with respect to timeliness metrics. The system is robust to the choice of a coding bucket size, so the number of packets combined in an encoded packet can vary around $K=10$, but any value $K>1$ results in better performance than the uncoded case $K=1$. These gains are observed for a wide range of channel conditions, with stable performance as long as the erasures are not extreme. The AoI takes small values as long as the system is kept away from extreme cases of low or high utilization, so there is a wide range, say $0.2<\rho<0.8$ where the system is forgiving to variations in arrival and service rates. The gains extend to general network topology. In fact, our results hold for a multi-hop multi-path scenario with the proper adjustment of the maximum feasible end-to-end rate such that adaptive coding  provides significant and robust gains with respect to timeliness in wireless networks. 

\begin{table}
\centering
\caption{Reduction in AAoI w.r.t. uncoded case ($K=1$)}
\begin{tabular}{ c|c|c|c|c } 
rate $r$ & $0.25$ & $0.50$ & $0.75$ & $1.0$ \\
 \hline
 \hline
$K=2, D=1$ & $1.87$ & $1.13$ & $0.88$ & $0.76$ \\
 \hline
$K=2, D=5$ & $4.95$ & $4.21$ & $3.99$ & $3.89$ \\
 \hline
$K=10, D=1$ & $3.91$ & $2.99$ & $2.77$ & $2.69$ \\
 \hline
$K=10, D=5$ & $13.95$ & $13.46$ & $13.38$ & $13.36$ 
\end{tabular}
\label{tab:gains}
\vspace{-0.1cm}
\end{table}

\section{Conclusion}
We studied the performance of adaptive coding of packet traffic with respect to timeliness metrics associated with the Age of Information (AoI). For a communication network modeled with an erasure channel and discrete time, we presented closed form expressions for the Average and Peak AoI as functions of a tunable parameter $K$ that defines the number of original packets to be coded together. While the benefits of network coding with respect to throughput and delay are well known and documented, this work has shown that AoI metrics may also be significantly improved by transmitting linear combinations of a few original packets. We observed that the AoI is decreasing for a large range of values for $K$, and noted that the biggest gain is obtained when departing from the (uncoded) case of $K=1$, so coding a small number ($K\leq 10$) packets together was demonstrated to greatly improve performance in systems that are sensitive to information timeliness. We showed that these AoI gains are robust to variations in $K$, feedback delay, and end-to-end rate that encapsulates channel and network topology effects.

\section*{Acknowledgement of Support and Disclaimer} 

\noindent (a) Contractor acknowledges Government's support in the publication of this paper. This material is based upon work funded by AFRL, under AFRL Contract No. \# FA8750-18-C-0190. (b) Any opinions, findings and conclusions or recommendations expressed in this material are those of the author(s) and do not necessarily reflect the views of AFRL.

\end{document}